# An Efficient Gradient Projection Method for Structural Topology Optimization

Zhi Zeng[†] and Fulei Ma[†]
zhizeng@mail.xidian.edu.cn; fuleima@xidian.edu.cn
School of Mechano-Electronics Engineering, Xidian University, Xi'an, Shaanxi
710071, China

**Abstract**

*This paper presents an efficient gradient projection-based method for structural topological optimization problems characterized by a nonlinear objective function which is minimized over a feasible region defined by bilateral bounds and a single linear equality constraint. The specialty of the constraints type, as well as heuristic engineering experiences are exploited to improve the scaling scheme, projection, and searching step. In detail, gradient clipping and a modified projection of searching direction under certain condition are utilized to facilitate the efficiency of the proposed method. Besides, an analytical solution is proposed to approximate this projection with negligible computation and memory costs. Furthermore, the calculation of searching steps is largely simplified. Benchmark problems, including the MBB, the force inverter mechanism, and the 3D cantilever beam are used to validate the effectiveness of the method. The proposed method is implemented in MATLAB which is open-sourced for educational usage.*

## 1. Introduction

Since the introduction of topology optimization to structural design, it has been successfully applied to many different types of structural design problems through various optimization schemes including density methods [1], boundary variation methods (like the level-set approach [2-4]), intelligent optimization methods (like the genetic evolutionary method [5]), etc. A comprehensive review can be found from Ref. [6-9].

One typical large class of topology optimization problems can be characterized by a nonlinear objective function which is minimized over a feasible region defined by bilateral bounds and a single linear equality constraint. In this paper, we focus on solving this class of optimal design problems which have a wide range of applications [3, 9]. The optimization problem can be mathematically expressed as:

---

[†] The authors contribute equally to this paper.

$$\min \quad O(\mathbf{x}) = \mathbf{l}^T \mathbf{U}$$

$$\text{s.t.} \quad \begin{cases} V(\mathbf{x})/V_0 = f \\ \mathbf{KU} = \mathbf{F}_o \\ x_{\min} \leq x_i \leq x_{\max} \end{cases} \tag{1}$$

$$\text{where} \quad \mathbf{x} = [x_1, x_2, \cdots, x_n]$$

in which **x** is the vector of design variables. Vector **l** in (1) takes different values for different problems. For example, $\mathbf{l} = \mathbf{F}_o$ for the minimum compliance design problem [3]. $n$ is the number of elements used to discretize the design domain. $O(\mathbf{x})$ is the objective function, **U** and $\mathbf{F}_o$ are the global displacements and the force vector respectively. **K** is the global stiffness matrix. $V(\mathbf{x})$ and $V_0$ are the material volume and design domain volume, respectively. Finally, $f$ is the prescribed volume fraction.

To solve the above list problem, different methods have been proposed and among which, the Solid Isotropic Material with Penalization for intermediate densities method (SIMP) [1] is considered the most effective material interpolation scheme thus has been widely implemented in industrial applications. The scheme is formulated as follows:

$$E_e(\mathbf{x}) = \mathbf{x}^r E_0 \tag{2}$$

where $E_e$ is the Youngs modulus, **x** is the vector of design variables which are constrained to [0, 1], $E_0$ is the stiffness of the material, $r$ is the penalization factor. Using this scheme, different approaches, including the Optimality Criteria (OC) method [10, 11], the Method of Moving Asymptote (MMA) [12], the Sequential Linear Programming (SLP) method [13, 14], and the Feasible Direction Method (FDM) [15-17], etc., have been proposed to solve the structural topology optimization problem and each approach has its desirable characteristics.

The OC method tends to convert the optimization to an equation-solving problem using the KT condition, which is attacked by iteratively approaching the fixed point [10]. The MMA method transforms the original problem into a series of localized strictly convex approximating subproblem which is solved by a dual method [12]. Similarly, the SLP method involves sequentially solving an approximate linear subproblem using the linearized objective and constraint functions [14]. The FDM, by utilizing the gradient at the current point, provides a feasible decent search direction and iteratively approach the optimal value. As a major feasible direction method, the gradient projection method (GPM) projects each step onto the feasible region [18].

To the best of our knowledge, comparing with the OC, MMA, and SLP, which have been extensively explored, relatively fewer investigations are performed on GPM for structural topology optimization. This may because that, people found using rudimentary methods like the raw GPM with no specialized scaling and projection scheme are less efficient [8]. While in a few current studies, people found that, given proper specialization, the GPM can be simple and effective for the current problem and has its advantages [15, 19]. Therefore, in the current study, we devote to improve this method and its variant in solving the problem defined in Eq. (1).

When applying the GPM to large scale problems like structural topology

optimization, the main difficulties we may face are list as follows:
- Effectively and efficiently obtain proper scaling for the gradient [18]
- Effectively and efficiently calculate the projection [18, 19]
- Effectively and efficiently obtain a proper searching step [18, 19]
- Numerical problems and others [20]

Although lots of research are devoted to alleviating these problems and details are discussed in the following section, there is still space for further improving the GPM for the current problem. Particularly, many effective engineering experiences in improving structural topology optimization are not fully exploited such as various effective sensitivity filters and post processing techniques [8, 15, 21, 22]. In a board sense, these techniques implicitly improve the scaling scheme and searching step. Whether we could propose a more effective method using these techniques is discussed in this paper, and some effective and efficient techniques are proposed.

The scope of the current study is given as follows. The problem statement is presented in Sec. 2. The proposed method is presented in Sec. 3. Sec. 4 provides three benchmark problems to validate the effectiveness of the proposed method. Concluding remarks are provided in Sec. 5.

## 2. Problem Statement

For the structural topology optimization problem, the differentiable cost function $O$ is minimized over a closed convex polyhedron $X \subset \mathbb{R}^n$. In the feasible direction method, one seeks for a feasible sequence $\mathbf{x} \subset X$ with an iteration of the form

$$\mathbf{x}_{k+1} = \mathbf{x}_k + \alpha_k \mathbf{d}_k \tag{3}$$

in which $\alpha_k \in \mathbb{R}^+$ and $\mathbf{x}_k + \alpha_k \mathbf{d}_k \in X$ for small enough $\alpha_k$ [23, 24]. (By stating small enough $\alpha_k$, we mean that the step size should be chosen properly, so that the new iterate belongs to the feasible region.) The subscript $i$ denotes the iteration number. To forestall any misunderstanding, we must state from the outset that the terminology used in the current study is consistent with those used in Ref. [24]. As one major feasible direction method [24], the gradient projection method has the form

$$\mathbf{x}_{k+1} = P_X \left( \mathbf{x}_k + \alpha_k \nabla f(\mathbf{x}_k) \right) \tag{4}$$

where $P_X(\cdot)$ denotes projection on $X$. The original gradient projection method has two significant drawbacks [24]. The first one is that its convergence is similar to the one of steepest descent, which is often slow. This problem is commonly alleviated through scaling, like the well-known projected Newton methods [25, 26]. The general form of such a scaling scheme is formulated as

$$\mathbf{x}_{k+1} = P_X \left( \mathbf{x}_k + \alpha_k \mathbf{D}_k \nabla f(\mathbf{x}_k) \right) \tag{5}$$

in which $\mathbf{D}_k$ is a scaling matrix. The second one is the projection operation may involve substantial overhead. To solve this problem, one prefers to take advantage of the specialty of the constraint, like bounds on $\mathbf{x}$ together with a single linear constraint [19, 27, 28].

In practice, another problem we must face is that the scale of the structural topology optimization is large, which results in much overhead in calculating $\mathbf{D}_k$ and the step size $\alpha_k$. Common remedies include using approximations to the Hessian matrix $(\nabla^2 f(\mathbf{x}_k))^{-1}$ [25-27, 29] and using constant or adaptive cyclic reusing of the Barzilai-Borwein step as the initial step size [19] instead of the line minimization rule.

Note that engineering experiences are always embedded in optimization algorithms to fully inspire their potentials. Such experiences include gradient modifications used in large scale non-linear optimization problems [30, 31], sensitivity filters [8, 22], and grey elements suppression techniques [15, 21]. In a board sense, these techniques implicitly improve the scaling scheme and searching step. While to the best of our knowledge, these techniques are commonly served as auxiliary preprocessing or postprocessing methods for structural topology optimization, and the main part of the optimization algorithms are not taken full advantages of these operations. Especially, whether we could propose a more effective (or simpler) scaling scheme, projection method, and searching step using these techniques are under-studied. The details will be discussed in the following section.

## 3. Efficiency Improvement

Gradient modification in structural topology optimization is not a new technique and often appears in the form of a sensitivity filter to avoid certain kinds of local minimum [22]. Other forms like magnitude modifications where the magnitude is modified by its squared root can also be seen for accelerating the process of solving certain problems [32]. We found that the effect of the latter technique is similar to a well-known technique called gradient clipping [31].

Gradient clipping, when used in large scale non-linear optimization, involves thresholding the gradient values elementwise if the gradient exceeded an expected range [31]. When the traditional gradient descent algorithm proposes to make a very large modification, the gradient clipping heuristic intervenes to reduce the modification to be small enough that it is less likely to go outside the region where the gradient indicates the direction of approximately steepest descent [31]. Most recent research gave a theoretical proof that under certain conditions, using gradient clipping, the converge speed may be faster than gradient descent with fixed step size [30].

For structural topology optimization, it is observed that small portions of entries in the matrix or tensor of the gradient have a much larger magnitude compared with the mean value of gradients. This phenomenon is similar to that encountered in Ref. [30]. Particularly in the early stages of optimization, these large values exist in the fragile parts along the force-loading path. By stating 'early stages', we mean that the design variable is not in the near neighbor of its optimal position. The existence of these large values restricts the step size of updating and thus hinders the optimization efficiency for gradient decedent methods.

In the current study, we propose a gradient clipping strategy for structural topology optimization as given in Eq. (6). The threshold is set empirically to be five times the mean value of the magnitude of gradients. (Too small a value means one takes no use of the gradient, a situation which is undesired. On the contrary, an extremely large threshold means one does not clip the gradient.) This modification can be viewed as a generalized non-linear scaling scheme for the gradient, which can accelerate the optimization in the early stages of the optimization for many problems as illustrated in Section 4.

$$(\nabla f(\mathbf{x}_k))_{mod} = \max(-T, \min(T, \nabla f(\mathbf{x}_k))) \qquad (6)$$

Grey elements suppression is a kind of post-processing [21]. The primary purpose of these techniques is to generate more distinct solid and void designs. Generally, the post-processed structure has a performance that is very close to the optimized structure [21]. Note that special problems that are extremely binarization sensitive out pass the investigating range of the current study and will not be considered here.

A generalized expression for thresholding based grey elements suppression technique is formulated as follows,

$$(x_i)_{mod} = \begin{cases} x_{max} & x_i \geq x_{max} - \Delta_1 \\ l + x_i & \text{otherwise} \\ x_{min} & x_i \leq x_{min} + \Delta_2 \end{cases} \qquad (7)$$

$$\text{where } l = (V - V_0) \Big/ \sum_{x_{min} + \Delta_1 \leq x_i \leq x_{max} + \Delta_2} 1$$

in which $\Delta_1$ and $\Delta_2$ are predefined thresholds and the shifting factor $l$ is used to guarantee the volume fraction. It can be seen that if $x_{max} - \Delta_1 = x_{min} + \Delta_2$, Eq. (7) is equivalent to thresholding based binarization. In this paper, we empirically set $\Delta_1 = \Delta_2$ to be a small constant no more than 0.3 or 0.75 times the volume fraction. This treatment is similar to the constraint thickness given in Ref. [17], which is used to eliminate numerical problems such as zigzag stepping and others in feasible direction methods.

Except for eliminating numerical problems, Eq. (7) leads to a gap between the largest or the smallest grey value in **x** to the nearest bound, namely $x_{min}$ or $x_{max}$. It is found that we could take advantage of this gap to simplify the calculation of the projection and step size. In detail, one can simplify the projection onto a convex polyhedron to that onto a simplicial cone when performing gradient projection. Note that this assumption is reasonable only when Eq. (7) is performed and the step size $\alpha$ is sufficiently small. Otherwise, one should use the method in Refs. [19, 33] instead. Different from calculating the projection onto the simplicial cone using the common technique given in Ref. [33], the current study provides an even simpler analytical solution that can be used to approximate the result considering the special structure of the constraint. The details are given in the following.

To make the explanation clearer, we modify the problem stated previously a bit by using an equivalent problem which considering the projection of the search direction **d** onto an **x**-dependent simplicial cone $C(\mathbf{x})$ as follows:

$$\begin{cases} d_i \geq 0 & \text{when } x_i = x_{min} \\ d_i \leq 0 & \text{when } x_i = x_{max} \\ \sum d_i = 0 \end{cases} \qquad (8)$$

The first two inequality constraints in Eq. (8) means that if the current **x** is on the boundary of $X$, the search direction should not point to the outside of $X$. The last equality constraint indicates that along the search direction, the volume of $\mathbf{x}$ does not change.

Let $P_{C(\mathbf{x})}(\mathbf{d})$ be the projection of $\mathbf{d}$ onto $C(\mathbf{x})$. Then, if we know all inequality constraints that are active on $P_{C(\mathbf{x})}(\mathbf{d})$. (By stating that one inequality constraint is active on a vector, we mean that the equality holds for this inequality constraint.) Then, all corresponding inequality constraints in Eq. (8) can be changed by equality constraints, and inequality constraints corresponding to inactive ones can be discarded. Then, $P_{C(\mathbf{x})}(\mathbf{d})$ can be calculated by projecting $\mathbf{d}$ on to the null space of the matrix representing all active constraints (including the last equality constraint in Eq. (8)) [34]. The first problem we have to address is how to calculate the null space and corresponding projection in a cheap way. This is because when a large-scale problem is encountered, even a single matrix multiplication is time (and memory) costing, not to mention the matrix decomposition involved in obtaining the null space [35, 36]. Fortunately, owing to the special structure of the constraints in Eq. (8), we found that there is an analytical solution to this problem. Details are proposed as follows.

By concatenating all normal vectors of all active constraints on $P_{C(\mathbf{x})}(\mathbf{d})$ (in the form of row vectors for easy calculation in MATLAB), we obtain a matrix $\mathbf{N} \in \mathbb{R}^n \times \mathbb{R}^{m+1}$. Note that given proper shuffling of the indices of the components of $\mathbf{x}$, the matrix $\mathbf{N}$ can always be written in the following form.

$$\mathbf{N} = \begin{bmatrix} \mathbf{N}_1 & \mathbf{N}_2 \\ \mathbf{0} & \mathbf{I} \end{bmatrix} = \begin{bmatrix} 1 & \cdots & 1 & 1 & \cdots & 1 \\ 0 & \cdots & 0 & 1 & \cdots & 0 \\ \vdots & \ddots & \vdots & \vdots & \ddots & \vdots \\ 0 & \cdots & 0 & 0 & \cdots & 1 \end{bmatrix} \qquad (9)$$

We denote the bases of the orthogonal complement of the subspace spanned by $\mathbf{N}$ as $\mathbf{M} \in \mathbb{R}^n \times \mathbb{R}^{n-m}$. Then, $\mathbf{M}$ can be expressed as follows.

$$\mathbf{M} = \begin{bmatrix} \mathbf{M}_1 \\ \mathbf{0} \end{bmatrix} \qquad (10)$$

where $\mathbf{M}_1$ is the matrix consisting of the bases of the orthogonal complement of the subspace spanned by $\mathbf{N}_1$. This can be easily verified by noting that $\mathbf{N}^T\mathbf{M} = \mathbf{0}$ and $\mathbf{M}^T\mathbf{M} = \mathbf{I}$. To calculate $\mathbf{M}_1$, we proposed an analytical expression as follows:

$$\mathbf{M}_1 = \begin{bmatrix} m & m & \cdots & m \\ 1+x & x & \cdots & x \\ x & 1+x & \cdots & x \\ \vdots & \vdots & \ddots & \vdots \\ x & x & \cdots & 1+x \end{bmatrix} \qquad (11)$$

where

$$m = -\frac{1}{\sqrt{n_1}}$$

$$x = \frac{-1 + \frac{1}{\sqrt{n_1}}}{n_1 - 1} \quad (12)$$

in which $n_1$ is the number of columns in $\mathbf{M}_1$. $\mathbf{M}_1$ can be further decomposed into a low-rank matrix and a sparse one, that is

$$\mathbf{M}_1 = \mathbf{A} + \mathbf{B} = x \begin{bmatrix} 1 & 1 & \cdots & 1 \\ 1 & 1 & \cdots & 1 \\ 1 & 1 & \cdots & 1 \\ \vdots & \vdots & \ddots & \vdots \\ 1 & 1 & \cdots & 1 \end{bmatrix} + \begin{bmatrix} m-x & m-x & \cdots & m-x \\ 1 & 0 & \cdots & 0 \\ 0 & 1 & \cdots & 0 \\ \vdots & \vdots & \ddots & \vdots \\ 0 & 0 & \cdots & 1 \end{bmatrix} \quad (13)$$

Denoting $\mathbf{d} = [\mathbf{d}_1 \ \mathbf{d}_2]^T$, the projection $P_{C(\mathbf{x})}(\mathbf{d})$ can be expressed by

$$\begin{aligned} P_{C(\mathbf{x})}(\mathbf{d})^T &= \mathbf{d}\,\mathbf{M}\mathbf{M}^T \\ &= [\mathbf{d}_1 \ \mathbf{d}_2] \begin{bmatrix} \mathbf{M}_1 \\ \mathbf{0} \end{bmatrix} [\mathbf{M}_1^T \ \mathbf{0}] \\ &= [\mathbf{d}_1 \mathbf{M}_1 \mathbf{M}_1^T \ \mathbf{0}] \\ &= [\mathbf{d}_1 \mathbf{A}\mathbf{A}^T + \mathbf{d}_1 \mathbf{A}\mathbf{B}^T + \mathbf{d}_1 \mathbf{B}\mathbf{A}^T + \mathbf{d}_1 \mathbf{B}\mathbf{B}^T \ \mathbf{0}] \end{aligned} \quad (14)$$

Suppose there are $n_1$ elements in $\mathbf{d}_1$, then

$$\mathbf{d}_1 \mathbf{A}\mathbf{A}^T = (n_1 - 1)x^2 \left[ \sum_{i=1}^{n_1} d_1^i \ \sum_{i=1}^{n_1} d_1^i \ \cdots \ \sum_{i=1}^{n_1} d_1^i \right]$$

$$\mathbf{d}_1 \mathbf{A}\mathbf{B}^T = x \left[ (n_1 - 1)(m - x) \sum_{i=1}^{n_1} d_1^i \ \sum_{i=1}^{n_1} d_1^i \ \cdots \ \sum_{i=1}^{n_1} d_1^i \right]$$

$$\mathbf{d}_1 \mathbf{B}\mathbf{A}^T = x \begin{bmatrix} \sum_{i=1}^{n_1} d_1^i + ((n_1 - 1)(m - x) - 1) d_1^1 \\ \vdots \\ \sum_{i=1}^{n_1} d_1^i + ((n_1 - 1)(m - x) - 1) d_1^1 \end{bmatrix}^T$$

$$\mathbf{d}_1 \mathbf{B}\mathbf{B}^T = \begin{bmatrix} (m-x)\sum_{i=1}^{n_1} d_1^i + ((n_1 - 1)(m-x)^2 - (m-x)) d_1^1 \\ (m-x) d_1^1 + d_1^2 \\ (m-x) d_1^1 + d_1^3 \\ \vdots \\ (m-x) d_1^1 + d_1^3 \end{bmatrix}^T \quad (15)$$

Note that the proposed method gives an explicit analytical expression for the null space projection and no matrix multiplication exists. Therefore, for large scale problems, the proposed method is efficient both in memory and time costs.

The next problem is to find out all active constraints on $P_{C(\mathbf{x})}(\mathbf{d})$. (Please be careful that they are distinct from the active constraints on $\mathbf{d}$.) This can be done by checking the redundancy of constraints one by one with limited overhead since the projection operation given in Eq. (14) is so cheap. While we find in our experiments that the set of all active constraints on $P_{C(\mathbf{x})}(\mathbf{d})$ can be well approximated in practical by the expansion method in Algorithm 1. For most problems in practice, the procedure given in Algorithm 1 can be accomplished within a few steps. In addition, observation of less than 3% redundant constraints existing after running Algorithm 1.

---
**Algorithm 1**

**Initially**, let $i = 0$ and $\mathbf{N}_0 = [1, 1, …, 1]$
**Do** calculate $P_{C(\mathbf{x})}(\mathbf{d})$ according to $\mathbf{N}_i$ using Eq. (14)
**While** $P_{C(\mathbf{x})}(\mathbf{d})$ does not satisfy the constraint (8)
  ➢ Concatenate all vectors of conflicting constraints to $\mathbf{N}_i$
  ➢ Increase $i$ by one
**Finally**, one has an approximation of $\mathbf{N}$ as well as the corresponding $P_{C(\mathbf{x})}(\mathbf{d})$

---

The final step is to calculate the optimal searching step $\alpha$. Using line minimization rule is impractical for the large-scale problem unless reusing of values for multiple iterations is adopted. A more practical choice is to use a constant step [37] or the BB step used in Ref. [19] and clip the step by a maximum value $\alpha_{max}$, so that $\mathbf{x}_k + \alpha_k \mathbf{d}_k$ is feasible. The maximum value is defined as:

$$\alpha_{\max} = \min_j (\alpha_j) \tag{16}$$

in which

$$\alpha_j = \begin{cases} \dfrac{x_{\max} - x_j}{d_j} & d_j > 0 \\ \dfrac{x_j - x_{\min}}{-d_j} & d_j < 0 \\ +\infty & d_j = 0 \end{cases} \tag{17}$$

An interesting finding in experiments is that the maximum step $\alpha_{max}$ itself is a decent choice for the step size which shows an adequate amount of decrease for each iteration. Experiences show that, although this is not the optimal step size, it does take a good balance between computation cost and optimization efficiency for many problems. A possible explanation for this phenomenon may be that the SIMP scheme diminishes the contribution of elements with intermediate densities (gray elements) to the total stiffness. The penalty factor steers the optimization solution to elements that are either solid black or void white [1].

## 4. Numerical Examples

Three benchmark examples including an MBB beam (minimum compliance design problem), a compliant force inverter mechanism (compliant mechanism design problem), and a 3D cantilever beam are provided in this section to demonstrate the performance of the proposed method.

**MBB**

The benchmark problem of finding the optimal material distribution, associated with the MBB beam, in terms of minimum compliance, with a constraint on the total amount of material, is presented in this section as an example.

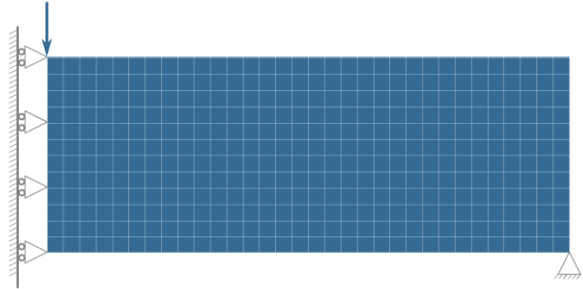

**Figure 1** Design domain of the MBB beam

In accordance with Ref. [38], the design domain, the boundary conditions, and the external load for the MBB beam are shown in Figure 1. The design domain is discretized with 60×20, 150×50, and 300×100 square elements, respectively. The volume fraction $f$ is set to 0.5. The Youngs modulus $E_0 = 1$ and the penalization factor is $r = 3$. Both the proposed method and the OC method are implemented in MATLAB and applied to this problem (the OC method is implemented using Andreassen's 88 lines of code in MATLAB [38]). As suggested by the authors, density and sensitivity filters with a radius of $r_{min} = 2.4$, 6 and 12, respectively, are used in the OC method to eliminate the numerical difficulties. Similarly, in our work, a Gaussian filter with a radius of $r_{min} = 1.1$, 2.0 and 4.0 for density and $r_{min} = 0.55$, 1.0 and 2.0 for sensitivity is used. In addition, we set $\Delta_1 = \Delta_2 = 0.3$. It should be noted that the filter sizes for both methods cannot be further reduced or local minima would emerge. All the experiments are executed on a personal computer with an Intel CORE i9 9900X processor, 128 GB memory, Windows 10 (64-bit), and MATLAB R2019b.

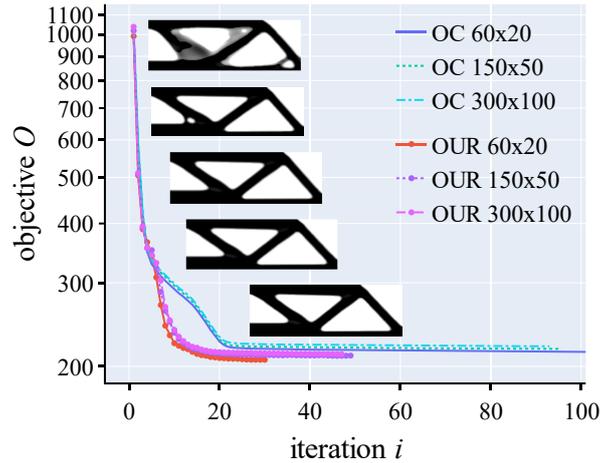

**Figure 2** Convergence curves for the MBB beam compliance minimization problem using two different methods

The normalized compliance as a function of iterations for both methods is plotted in Figure 2. Several conclusions can be drawn from this example. First of all, both methods almost converge in less than 30 iterations, which shows that they can efficiently converge to an optimum design from a uniform grey starting guess. Second, for both methods, a steep drop is observed at the initial several steps and the convergence speed of the OC method slows down after that. Results show that the proposed method can converge to a smaller normalized compliance value (approximately 3% improvement on average) with fewer iterations (approximately 60% improvement on average). Finally, a larger number of grey elements exist in the layout of the OC method, which shows that the proposed method is less likely to converge to chessboard patterns with smaller filters. The optimized designs of the MBB beam and corresponding compliance $O$ obtained using different methods with different refinements are illustrated in Figure 3. The results demonstrate that, given proper filter settings, different refinements do not lead to different topologies.

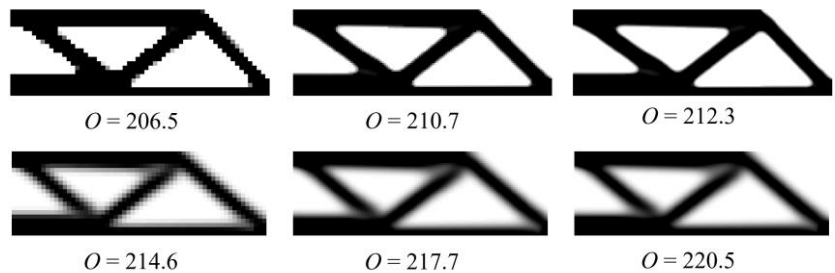

**Figure 3** Optimized design of the MBB beam and corresponding compliance $O$ obtained with the variant of the 88-line code (bottom) and the proposed one (top). A mesh with 60×20 elements (left), 150×50 elements (middle), and 300×100 elements (right) has been used.

## Gripper

To demonstrate the effectiveness of the proposed method in dealing with compliant mechanisms, a benchmark example, associated with the compliant force converter mechanism, is provided.

Figure 4 depicts the design domain of a compliant force inverter mechanism with single input-output behavior. The inverter outputs a displacement in the opposite direction to the actuating force. The fixed bound area and the external force $F_i$ are denoted in the figure.

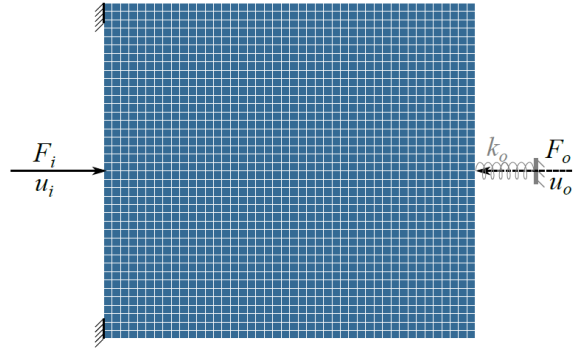

**Figure 4** Design domain and boundary conditions of the compliant force inverter mechanism

The design domain is discretized with 100×100 square elements. The volume fraction $f$ is set to 0.2. An artificial spring with stiffness $k_o$ is attached to the output port and a dummy load $F_o$ is applied at the output port, which is expected to produce a horizontal displacement $u_o$ to the left. The goal is to maximize the geometric advantage ( $G_A = u_o = u_i$ ) of the mechanisms. The objective function used in this paper is formulated as [39]:

$$\begin{cases} O = \max(G_A) \\ G_A = \dfrac{F_o \Delta_{21}}{F_o \Delta_{11} + \Delta_{11}\Delta_{22}k_o - \Delta_{21}\Delta_{12}k_o} \\ \Delta_{ij} = \dfrac{\mathbf{u}_i \mathbf{K}(\mathbf{x})\mathbf{u}_j}{F_i} \quad (i, j = 1, 2) \end{cases} \tag{18}$$

where $\mathbf{K}(\mathbf{x})$ is the global stiffness matrix of the structure. $\mathbf{u}_1$ and $\mathbf{u}_2$ represent the displacement fields of structures when only $F_i$ or the dummy load $F_o$ is applied, respectively. For the proposed method, a close operation [40] with a radius of $r_{min} = 1$ for density, $r_{min} = 1.84$ for sensitivity, and $\Delta_1 = \Delta_2 = 0.15$ is utilized.

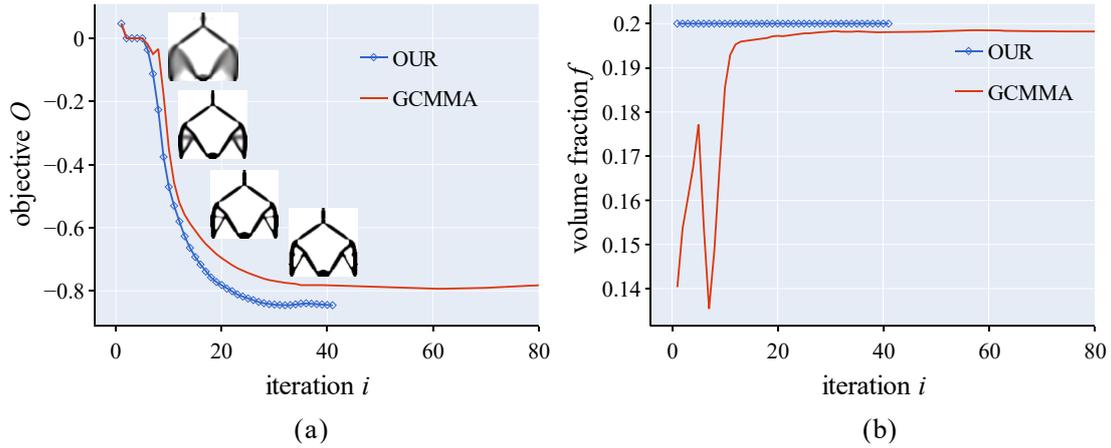

**Figure 5** Convergence curves for the complaint force inverter mechanism design problem with the $G_A$ set as the objective function. (a) The changes of objective functions; (b) The changes of volume fractions.

The $G_A$ as a function of iterations for the force inverter mechanism is plotted in Fig. 5. The iteration history and the associated layouts are plotted in the figure. To compare the performance of the proposed method with that of the GCMMA, the iteration history of GCMMA is also plotted.

We can find from the figure that both methods almost converge in less than 40 iterations, which shows that both approaches can efficiently locate a good design from a uniform grey starting guess. The proposed method shows a faster convergence speed (approximately 50% improvement) and converges to a smaller normalized compliance value (approximately 7% improvement) which shows that the proposed method is able to deal with compliant mechanisms optimization problems. The proposed method is more likely to converges to a binarization layout. While a larger number of grey elements exist in the layout of the MMA method. It should be noted that, though the proposed method shows a better quality in the example, it does not mean that this method is better than MMA for every case since multiple hyperparameters are included in both methods. The tuning of these parameters can also lead to an improvement. The optimized designs of the mechanism obtained using different methods are illustrated in Figure 6.

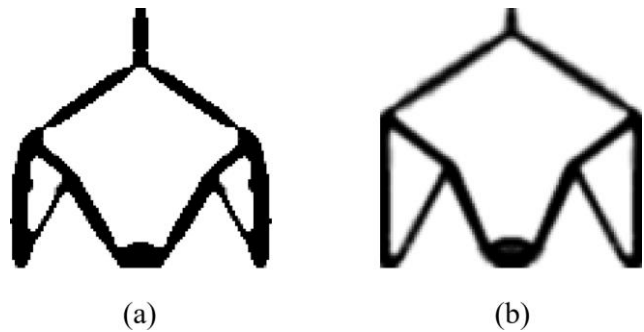

**Figure 6** Optimum topologies for the inverter mechanism (a) Layout of the proposed method (b) Layout of GCMMA

It should also be noted that the used algorithm is identical to the one used in the MBB example. The difference mainly lies in the objective function, which means that the proposed method can be applied to a different type of problem with less modification.

**3D Cantilever Beam**

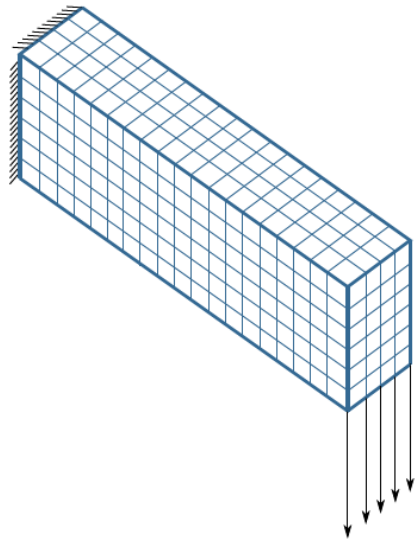

**Figure 7** Design domain of a 3D cantilever beam

To show the effectiveness of the proposed method in dealing with 3D problems, a cantilever beam problem is provided in this section. The design domain of the cantilever beam is illustrated in Fig. 7. The prismatic beam is fully constrained in one end and a unit distributed vertical load is applied downwards on the lower free edge of the other end. The dimension of the computational domain is 60×20×10.

Different methods including the OC method, the GCMMA and the proposed one are used to solve this problem. The settings for the OC method are identical to those used in Ref. [41]. The settings for the GCMMA are identical to those used in Ref. [42]. The generalized compliance as a function of iteration for the three methods is plotted in Fig. 8.

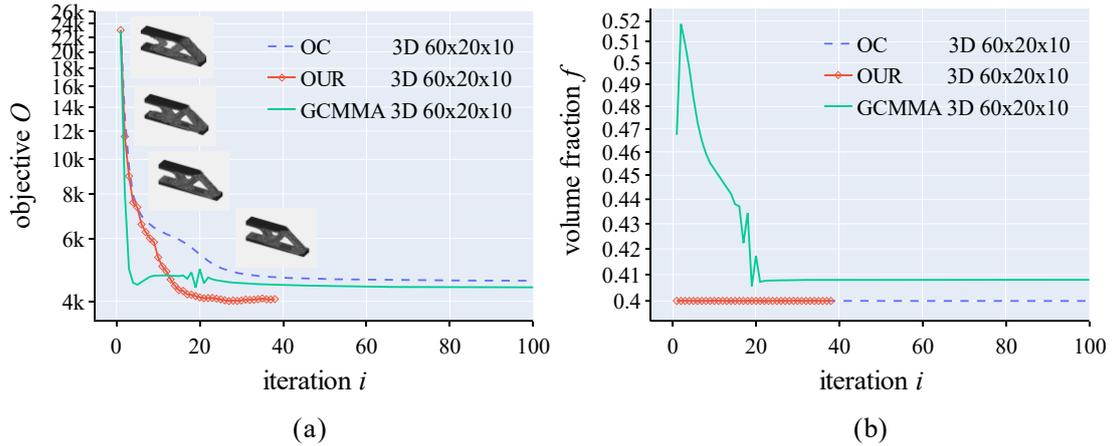

(a)                                   (b)

**Figure 8** Convergence curves for the 3D cantilever beam using the OC, GCMMA, and the proposed method. (a) The changes of objective functions; (b) The changes of volume fractions.

We can find from the results that all three methods converge rapidly in the first several steps. While, the OC method slows down quickly and converges to a relatively larger magnitude of generalized compliance. The proposed method gains approximately 7% improvement in the magnitude of generalized compliance comparing with that of the GCMMA and 11% improvement comparing with that of the OC. The steepest convergence rate is observed for GCMMA from Figure 8. The reason can be attributed to the fact that the material fraction constraint is not strictly satisfied at the initial stage for GCMMA (in other words, more material is contained in the mechanism). The optimized designs at different iterations for the proposed method are also illustrated in Fig. 8.

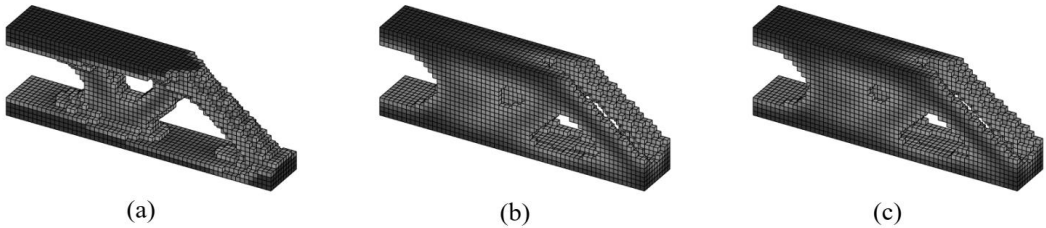

(a)                          (b)                          (c)

**Figure 9** Optimum topologies for the 3D cantilever beam with a mesh of 60×20×10. (a) The proposed one; (b) The OC method; (c) The GCMMA.

The final designs of the 3D cantilever beam using different methods are illustrated in Fig. 9. The final layout obtained using the OC and the GCMMA are similar to each other. While the topology obtained by the proposed method (shown in Fig. 9(a)) is quite different from and more compact comparing with Fig. 9(b) and Fig. 9(c). The results show that the proposed method is more likely to locate an optimum result.

The efficiency of a method is affected by two things. One is the convergence rate, the other is the computation costs. Since the projection in the current study has an analytical formulation and is dimension irrelative, the computation is fast. For all methods, the time costs for each iteration depend on both the FEA process and an updating procedure. The updating time costs for different methods in each iteration is plotting in Fig. 10. The results show that the updating cost of the proposed method is negligible compared to the other two methods. Note that as the scale of the problem increases, the time costs due to the FEA process is dominant. Nevertheless, the proposed method would find its advantage when the time and memory costs for the updating procedure matters.

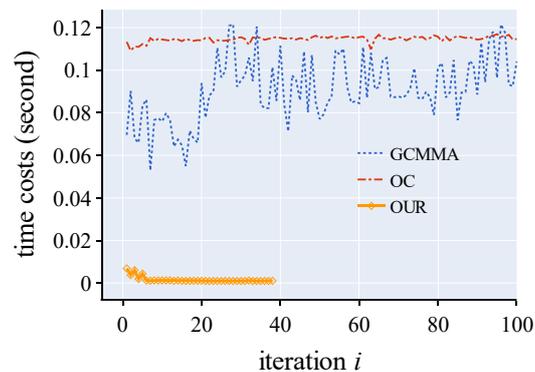

**Figure 10** Time costs of the updating process (exclude the FEA process) for each iteration of the three methods

## 5. Conclusion and Discussion

In this work, we proposed a modified gradient projection method with improved efficiency and neglectable loss of accuracy for structural topology optimization problems. Through gradient clipping, as well as the modified projection, the efficiency of the original gradient projection method has been greatly improved. It should be noted that since the projection is approximated by an analytical expression, the calculation involves negligible computation and memory costs. In addition, the determination of searching steps can be simplified accordingly.

Benchmark problems, including the MBB, force inverter mechanism, and 3D beam are analyzed using the proposed method, the results validate the effectiveness of the method. The method is also implemented in MATLAB and open-sourced for educational usage. It is recommended that the readers try our code for a better understanding of the proposed method.

It should be noted that adaptive filters and hybrid method may improve the efficiency to a certain extent. But this part is out of the scope thus will be considered in the future work.

## 6. Replication of results

The method proposed in this paper is implemented in MATLAB and open-sourced on GitHub for educational usage. (https://github.com/zengzhi2015/EGP) For commercial usage, please contact the authors.

## 7. Compliance with ethical standards

The authors declare that they have no conflict of interest.

**Acknowledgment**

The authors gratefully acknowledge the financial support from the National Natural Science Foundation of China under Grant No. 51805397 and No. 61805185 and the Open Fund of State Key Laboratory of Robotics and System (HIT) under Grant No. SKLRS-2019-KF-07